\documentclass[preprint,superscriptaddress]{revtex4}

\usepackage{graphicx}
\usepackage{array}
\usepackage{amssymb}
\usepackage{amsfonts}
\usepackage{amsmath}
\usepackage{mathrsfs}
\usepackage{color}
\usepackage{booktabs}
\usepackage{threeparttable}
\usepackage{multirow}
\usepackage{subfigure}
\usepackage{times}
\usepackage{epsfig}
\usepackage{threeparttable}
\usepackage{chngpage}
\usepackage{float}

\usepackage{color}

\makeatletter
\renewcommand\@biblabel[1]{#1.}
\makeatother

\begin{document}

\title{Exact Controllability of Complex Networks}

\author{Zhengzhong Yuan}
\affiliation{School of Systems Science, Beijing Normal University,
Beijing, 100875, P. R. China}

\author{Chen Zhao}
\affiliation{School of Systems Science, Beijing Normal University,
Beijing, 100875, P. R. China}

\author{Zengru Di}
\affiliation{School of Systems Science, Beijing Normal University,
Beijing, 100875, P. R. China}

\author{Wen-Xu Wang}\email{wenxuwang@bnu.edu.cn}
\affiliation{School of Systems Science, Beijing Normal University,
Beijing, 100875, P. R. China}
\affiliation{School of Electrical, Computer and Energy Engineering, Arizona
State University, Tempe, Arizona 85287, USA}

\author{Ying-Cheng Lai}
\affiliation{School of Electrical, Computer and Energy Engineering, Arizona
State University, Tempe, Arizona 85287, USA}
\affiliation{Department of Physics, Arizona State University, Tempe,
Arizona 85287, USA}

\begin{abstract}
Controlling complex networks is of paramount importance in science
and engineering. Despite the recent development of structural-controllability
theory, we continue to lack a framework
to control undirected complex networks, especially given link
weights. Here we introduce an exact-controllability paradigm based on
the maximum multiplicity to identify the minimum set of
driver nodes required to achieve full control of networks with
arbitrary structures and link-weight distributions.
The framework reproduces the structural controllability of directed
networks characterized by structural matrices. We explore the controllability of
a large number of real and model networks, finding that
dense networks with identical weights are difficult to be controlled.
An efficient
and accurate tool is offered to assess the controllability of large sparse
and dense networks.
The exact-controllability framework enables a comprehensive understanding of
the impact of network properties on controllability, a fundamental problem
towards our ultimate control of complex systems.
\end{abstract}

\maketitle

One of the most challenging problems in modern network science and engineering
is controlling complex networks. While great effort has been devoted to
understanding the interplay between complex networks and dynamical processes
taking place on them
\cite{AB:2002,Newman:2003,BLMCH:2006,Caldarelli:2007,Fortunato:2010}
in various natural and technological systems
\cite{WS:1998,BA:1999,AJB:1999,ASBS:2000,AJB:2000,CEBH:2000,JMBO:2001,PV:2001,NWS:2002,
PDFV:2005,NABV:2010}, the control of complex dynamical networks remains to be
an outstanding problem. Generally, because of the ubiquity of
nonlinearity in nature, one must consider the control of complex networked
systems with nonlinear dynamics. However, at present there is no general
framework to address this problem because of the extremely complicated interplay
between network topology and nonlinear dynamical processes, despite the
development of nonlinear control
\cite{SL:book,WC:2002,WS:2005,SBGC:2007,YCL:2009,RME:2009} in
certain particular situations such as consensus~\cite{EMCCB:2012},
communication~\cite{KMT:1998,CLCD:2007}, traffic~\cite{Srikant:book} and
device networks~\cite{Luenberger:book,SL:book}. To ultimately develop
a framework to control complex and nonlinear networks, a necessary and
fundamental step is to investigate the controllability of complex networks
with linear dynamics. There exist well developed theoretical frameworks
of controllability for linear dynamical systems in the traditional field
of engineering control \cite{Kalman:1963,Lin:1974,SP:1976,RW:1997,Sontag:book}.
However, significant challenges arise when applying the traditional controllability
framework to complex networks~\cite{LH:2007} due to the difficulty to determine
the minimum number of controllers.
A ground-breaking recent contribution was made by Liu et al. \cite{LSB:2011}
who developed a minimum input theory to efficiently characterize the
structural controllability of directed networks, allowing a minimum
set of driver nodes to be identified to achieve full control. In
particular, the structural controllability of a directed network can be
mapped into the problem of maximum matching~\cite{HK:1973,ZOY:2003,ZM:2006},
where external control is necessary for every unmatched node.
The structural-controllability framework also allows several basic
issues to be addressed, such as linear edge dynamics~\cite{NV:2012},
lower and upper bounds of energy required for control~\cite{YRLLL:2012},
control centrality~\cite{LSB:2012}, and optimization~\cite{WNLG:2012}.

Although the structural-controllability theory offers a general tool
to control directed networks, a universal framework for exploring
the controllability of complex networks with arbitrary structures and
configurations of link weights is lacking. Mathematically, the framework
of structural controllability is applicable to directed networks characterized by
structural matrices, in which all links are represented by independent free
parameters~\cite{LSB:2011}. This requirement may be violated if exact link weights are
given, motivating us to pursue an alternative framework beyond the
structural-controllability theory. For undirected networks, the symmetric characteristic
of the network matrix accounts for the violation of the assumption
of structural matrix, even with random weights. Thus we continue to lack
a reliable tool to measure the controllability of undirected networks.
For some practical issue towards achieving actual control, such as predicting
control energy with given link weights, necessary and sufficient conditions to
ensure full control are the prerequisite. Taken together, we need a more general and
accurate framework to study the controllability of complex networks.

In this paper, we develop an exact-controllability framework as an
alternative to the structural-controllability framework, which offers
a universal tool to treat the controllability of complex networks with
arbitrary structures and link weights, including directed, undirected, weighted
and unweighted networks with or without self-loops. Structural controllability
can be reproduced by our framework for structural matrix that can be ensured by assigning
random weights to directed links. In particular, based on the
Popov-Belevitch-Hautus (PBH) rank condition~\cite{PBH} that is equivalent to
the Kalman rank condition~\cite{Kalman:1963}, we prove that the minimum number
of independent driver nodes or external controllers is equal to the maximum geometric
multiplicity of all eigenvalues of the network matrix. If the network
matrix is diagonalizable, e.g., as for undirected networks, controllability is
simply determined by the maximum algebraic multiplicity of all eigenvalues.
That is, the minimum number of inputs is determined by the dimension of
eigenvectors for arbitrary networks and, for symmetric networks, this number
is nothing but the eigenvalue degeneracy. For simple regular networks, their
exact controllability can be calculated analytically. For more complicated
model networks and many real-world weighted networks with distinct node-degree
distributions, the exact controllability can be efficiently assessed by numerical computations.
The minimum set of driver nodes can be identified by
elementary transformations based on the exact-controllability framework.
A systematic comparison study indicates that the results from
our exact-controllability theory are consistent with those in Ref.~\cite{LSB:2011}
for cases where both frameworks are applicable. Application of our framework
also reveals a number of phenomena that cannot be uncovered by the
structural-controllability framework. For example, we find that for
random~\cite{ER:1960} and small-world networks~\cite{WS:1998,NW:1999} with
identical link weights, the measure of controllability is a non-monotonic
function of link density with largest controllability occurring in the
intermediate region. For highly sparse or dense networks, the former being
ubiquitous in real-world systems, the exact-controllability theory can be
greatly simplified, leading to an efficient computational paradigm
in terms solely of the rank of matrix.
\\

\noindent
\textbf{\textsf{Results}}
\\

\noindent
{\bf Exact controllability measurement of complex networks}\\

A necessary step towards ultimately controlling nonlinear network systems
is to fully understand the controllability of complex networks with linear
dynamics. Consider a network of $N$ nodes described by the following set of ordinary
differential equations:
\begin{equation} \label{eq:linearSys}
\dot{\mathbf{x}}=A\mathbf{x}+B\mathbf{u},
\end{equation}
where the vector $\mathbf{x} = (x_1,\cdots,x_N)^\text{T}$ stands for the states of nodes,
$A\in R^{N\times N}$ denotes the coupling matrix of the system, in which $a_{ij}$ represents
the weight of a directed link from node $j$ to $i$ (for undirected networks, $a_{ij}=a_{ji}$).
$\mathbf{u}$ is
the vector of $m$ controllers: $\mathbf{u}=(u_1,u_2,\cdots,u_m)^\text{T}$, and $B$
is the $N\times m$ control matrix. The classic Kalman rank condition
\cite{Kalman:1963,Rugh:book} stipulates that system (\ref{eq:linearSys}) can be
controlled from any initial state to any final state in finite time if and only if
$\text{rank}(C_1)=\text{rank}[B,AB,A^2B,...,A^{N-1}B]=N$. The standard way
to address the controllability problem is to find a suitable control matrix
$B$ consisting of a minimum number of driver nodes so as to satisfy the Kalman rank
condition. However, a practical difficulty is that there are $2^N$ possible combinations
of driver nodes. The notion of structural controllability was then introduced
by Lin \cite{Lin:1974} and was recently adopted to directed complex networks
\cite{LSB:2011}. Relying on the maximum matching algorithm, the
structural-controllability framework developed by Liu et al. enables efficiently
identifying a minimum set of unmatched nodes that need to be controlled to
achieve full control of the system. Thus, the unmatched nodes are the driver nodes,
the number of which determines the controllability of the network.
The computation complexity of finding $B$ based on the efficient
maximum matching algorithm can therefore be significantly
reduced. In general, the structural-controllability theory is
valid for directed networks associated with structural matrices in which all
links are represented by independent free parameters. This limitation
prompts us to develop an exact-controllability framework for arbitrary
network structures and link weights.

To formulate a theoretical framework of exact controllability, we need a new
starting point. Our theory is based on the PBH rank condition~\cite{PBH,Rugh:book,AM:book,Sontag:book}
from control theory, according to which system (\ref{eq:linearSys}) is fully
controllable if and only if
\begin{equation} \label{eq:PBH_origin}
\text{rank}[cI_N-A,B]=N
\end{equation}
is satisfied for any complex number $c$, where $I_N$ is the identity matrix of
dimension $N$. It can be proved that, if and only if any eigenvalue $\lambda$
belonging to matrix $A$ satisfies Eq.~(\ref{eq:PBH_origin}), full control can
be ensured~\cite{PBH,Rugh:book,AM:book,Sontag:book}. For system (\ref{eq:linearSys}), there are
many possible control matrices $B$ that satisfy the controllable condition.
The central goal is to find a set of $B$ corresponding to the minimum number $N_\text{D}$ of
independent drivers or controllers required to control the whole network. In
general, $N_\text{D}$ is defined in terms of $B$ as
$N_\text{D}=\min \{\text{rank}(B)\}$,
as in the structural-controllability based formulation \cite{LSB:2011}. The PBH
condition has been employed recently to study the controllability of leader-follower
systems~\cite{LFsys} and regular graphs~\cite{PN:2012,NP:2012}.

We develop a general theory to calculate $N_\text{D}$ based on the PBH rank condition (see Method and Supplementary Note~1).
It can be proven that for arbitrary network structures and link weights, say arbitrary
matrix $A$, the minimum number $N_\text{D}$ of controllers or drivers is determined by the
maximum geometric multiplicity $\mu(\lambda_i)$ of the eigenvalue $\lambda_i$ of $A$:
\begin{equation} \label{eq:geometric}
N_\text{D}=\max_i\{\mu(\lambda_i)\},
\end{equation}
where $\mu(\lambda_i)=\dim V_{\lambda_i}=N-\text{rank}(\lambda_i I_N-A)$ and
$\lambda_i$ $(i=1,...,l)$ represents the distinct eigenvalues of $A$.
For undirected networks with arbitrary link weights, $N_\text{D}$ is determined by
the maximum algebraic multiplicity $\delta(\lambda_i)$ of $\lambda_i$:
\begin{equation} \label{eq:algebra}
N_\text{D}=\max_i\{\delta(\lambda_i)\}.
\end{equation}
where $\delta(\lambda_i)$ is also the eigenvalue degeneracy of matrix $A$. We need to
stress that Eq.~(\ref{eq:geometric}) is rigorous and generally valid without any limitations for
matrix $A$ and Eq.~(\ref{eq:algebra}) is proven to be valid for general undirected networks
with diagonalizable matrix $A$.

To compute the geometric and algebraic multiplicities for large networks is
computationally demanding. However, the task can be greatly simplified if the network
is either sparse or dense. In fact, complex networks arising in real-world applications
are typically sparse \cite{AB:2002}. For a large sparse network, in which the number of
links scales with $N$ in the limit of large $N$~\cite{Newman:2004}, with a small fraction
of self-loops, $N_\text{D}$ is simply determined by the rank of $A$:
\begin{equation} \label{eq:efficient1}
N_\text{D} = \max\{1,N-\text{rank}(A)\}.
\end{equation}
For a densely connected network with identical link weights $w$, in which the zeros in $A$ scales
with $N$ in the limit of large $N$, with a small fraction of self-loops, we have
\begin{equation} \label{eq:efficient2}
N_\text{D} = \max\{1,N-\text{rank}(wI_N+A)\}.
\end{equation}
Relying on the simplified formulas (\ref{eq:efficient1}) and (\ref{eq:efficient2}),
$N_\text{D}$ can be evaluated by computing the rank of the network matrix $A$ or $wI_N +A$
in an efficient manner. For undirected sparse networks,
LU decomposition with computation complexity $O(N^{2.376})$~\cite{LU}
can yield reliable results of $\text{rank}(A)$. For other types of networks, SVD
method with $O(N^3)$~\cite{SVD} can give accurate results of $\text{rank}(A)$ as well as the
eigenvalues of $A$.

The measure of controllability
$n_\text{D}$, or simply controllability, of a network can be defined as the ratio of $N_\text{D}$
to the network size $N$
\cite{LSB:2011}:
\begin{equation}
n_\text{D} = N_\text{D} /N.
\end{equation}

\begin{figure}
\begin{center}
\epsfig{figure=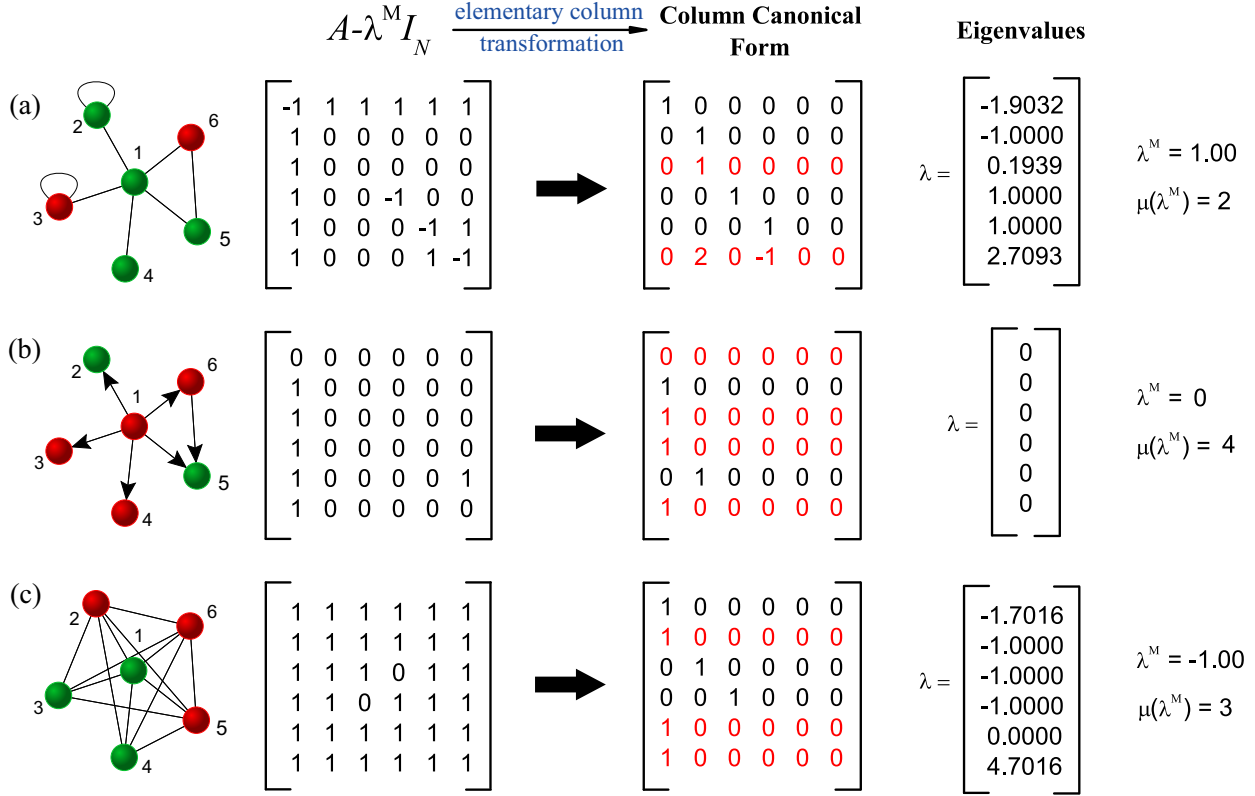,width=\linewidth}
\caption{ {\bf Illustration of the exact-controllability framework to identify
a minimum set of drivers}. (a) a simple undirected network with self-loops, (b)
a simple directed network and (c) an undirected network with dense connections.
The matrix $A-\lambda^\text{M} I_N$, the column canonical form of matrix $A-\lambda^\text{M} I_N$
by the elementary column transformation, the eigenvalues $\lambda$ and the eigenvalue
$\lambda^\text{M}$ corresponding to the maximum geometric multiplicity $\mu(\lambda^\text{M})$ of
$A$  are given for each simple network. The rows that are linearly dependent
on others in the column canonical form are marked by red. The nodes
correspond to them are the drivers that are marked by red as well in the networks.
For the undirected networks in (a) and (c), $\mu(\lambda^\text{M})$ is equal to the maximum
algebraic multiplicity, i.e, the multiplicity of $\lambda^\text{M}$.
The configuration of drivers is not unique as it relies on the elementary column
transformation, but the number of drivers is fixed and determined by the
maximum geometric multiplicity $\mu(\lambda^\text{M})$ of matrix $A$.}
\label{fig:illus}
\end{center}
\end{figure}

\noindent
{\bf Identify driver nodes}
\\

We offer a general approach to identifying a minimum set of
driver nodes that need to be controlled by relying on the PBH rank condition
(\ref{eq:PBH_origin}). According to the exact-controllability theory, $N_\text{D}$
is determined by the maximum geometric multiplicity $\mu(\lambda^\text{M})$ arising at the eigenvalue
$\lambda^\text{M}$. Thus, the control matrix $B$ to ensure full control
should satisfy the condition (\ref{eq:PBH_origin})
by substituting $\lambda^\text{M}$ for the complex number $c$, as follows:
\begin{equation}
\text{rank}[\lambda^\text{M}I_N-A,B] = N.
\label{eq:drivers}
\end{equation}
The question becomes how to find the minimum set of drivers identified in $B$ to
satisfy Eq.~(\ref{eq:drivers}). We note that the rank of the matrix $[\lambda^\text{M}I_N-A,B]$
is contributed by the number of linearly-independent rows. In this regard,
we implement elementary column transformation on the matrix $\lambda^\text{M}I_N-A$ (or $A-\lambda^\text{M} I_N$), which
yields a set of linearly-dependent rows that violate the full rank condition (\ref{eq:drivers}).
The controllers located via $B$ should be imposed on the identified rows to eliminate
all linear correlations to ensure condition (\ref{eq:drivers}). The nodes corresponding to
the linearly-dependent rows are the drivers with number
$N- \text{rank}(\lambda^\text{M} I_N - A)$, which is nothing but the maximum geometric multiplicity $\mu(\lambda^\text{M})$.
Note that each column in $B$ can at most eliminate one linear correlation, such that the minimum
number of columns of $B$, say, $\min\{ \text{rank}(B)\}$ is the same as the number $\mu(\lambda^\text{M})$ of drivers,
regaining our exact-controllability theory.

To illustrate the method explicitly, we present three typical simple examples, as exemplified in
Fig.~\ref{fig:illus}. For each graph, we first compute the eigenvalues $\lambda$ of the matrix $A$ and
their geometric multiplicity $\mu(\lambda)$ to find the eigenvalue $\lambda^\text{M}$ corresponding
to the maximum geometric multiplicity $\mu(\lambda^\text{M})$. The matrix $A-\lambda^\text{M} I_N$
associated with $\lambda^\text{M}$ can then be established so as to identify linear correlations. We perform
elementary column transformation on $A-\lambda^\text{M} I_N$ to obtain its column canonical form that
reveals the linear dependence among rows. The rows that are linearly-dependent on other rows correspond
to the driver nodes that need to be controlled to maintain full control. Note that the configuration
of drivers is not unique as it depends on the order of implementing elementary transformation and many possible choices
of linearly-dependent rows. Nevertheless, the minimum number of drivers is fixed and determined by
$\mu(\lambda^\text{M})$, analogous to the structural-controllability theory. The approach of
finding drivers is rigorous, ensured by the PBH rank condition, the maximum multiplicity theory and
the column canonical form associated with matrix rank. Thus, the method is applicable to arbitrary networks,
including networks described by structural matrices. We have tested its performance on structural matrix
by assigning each link a random parameter,
which shows excellent agreement with the result from the structural-controllability framework (Supplementary Fig.~S1 and Note~2),
but with different computational efficiency. The maximum matching algorithm with computation complexity $O(N^{1/2}L)$
($L$ denotes the number of links) outperforms our
elementary-transformation-based algorithm with $O(N^2(\log N)^2)$~\cite{GaussEle} for dealing with structural matrix.

In general, we have developed a framework to study the exact controllability of
arbitrary complex networks by means of the maximum multiplicity of network
matrix to quantifying the minimum number of drivers and the elementary column transformation
to detect all drivers. Applying the framework to model and real networks will
gain deep insight into the significant problem of network control.
\\

\begin{table}
\begin{ruledtabular}
\begin{center}
\caption{{\bf Eigenvalues and minimum number of driver nodes of regular unweighted and undirected
graphs}. $N_\text{D}^\text{MMT}$ denotes the minimum number of drivers calculated from
the maximum algebraic multiplicity. $q=1,2, \cdots, N$ and the algebraic multiplicity of eigenvalues is
indicated in ``()'' for star and fully connected networks.}
\label{tb-1}
\begin{tabular}{lcc}
Network & Eigenvalue & $N_\text{D}^\text{MMT}$\\
\hline
Chain & $2\cos{\frac{q\pi}{N+1}}$ & $1$ \\
Ring Network & $2\cos{\frac{2\pi(q-1)}{N}},\lambda_q=\lambda_{N-q+2}$& $2$\\
Star Network& $0(N-2),\pm{\sqrt{N-1}}(1)$  & $N-2$\\
Fully Connected Network&$N-1(1),-1(N-1)$&$N-1$\\
\end{tabular}
\end{center}
\end{ruledtabular}
\end{table}

\noindent
{\bf Exact controllability of model and real networks}
\\

We now present controllability measure $n_\text{D}$ of a large number of model networks with
identical- and random-weight distributions and real-world unweighted and
weighted networks from the exact-controllability theory and the structural-controllability
theory in cases where it is applicable. In many
real situations, we are unable to have exact link weights due to the
measurement noise and the nonlinear effects. Thus, it is reasonable to assume
random-weight distributions rather than identical weights when dealing with real networks.
However, in some man-made networks, identical weights or a small number of
different weights are possible, prompting us to consider the controllability
of unweighted networks. Furthermore, revealing the effect of network topology
on a variety of dynamics is of paramount theoretical importance in network
science and has been given tremendous interest~\cite{AB:2002,Newman:2003,BLMCH:2006}.
As the central issue related to network dynamics, the influence of topology to
controllability deserves to be fully understood in unweighted networks without
complications from other factors. Considered together, we study $n_\text{D}$ with respect to
unweighted networks (identical weights) and networks with random weights.

For undirected regular graphs with identical weights,
$N_\text{D}$ and the relevant eigenvalues can be calculated precisely, as listed in
Table~\ref{tb-1}. We see that the chain and ring graphs can be controlled by one and two
controllers, respectively. However, for the star and fully connected networks,
almost all nodes need to be independently controlled. The results of chain
and star networks are consistent with those from the structural-controllability
theory~\cite{LSB:2011}, but the results of ring and fully-connected networks
are different (Supplementary Note~3). This difference, as will be explained below, indicates the
need of the exact-controllability framework.

\begin{figure}
\begin{center}
\epsfig{figure=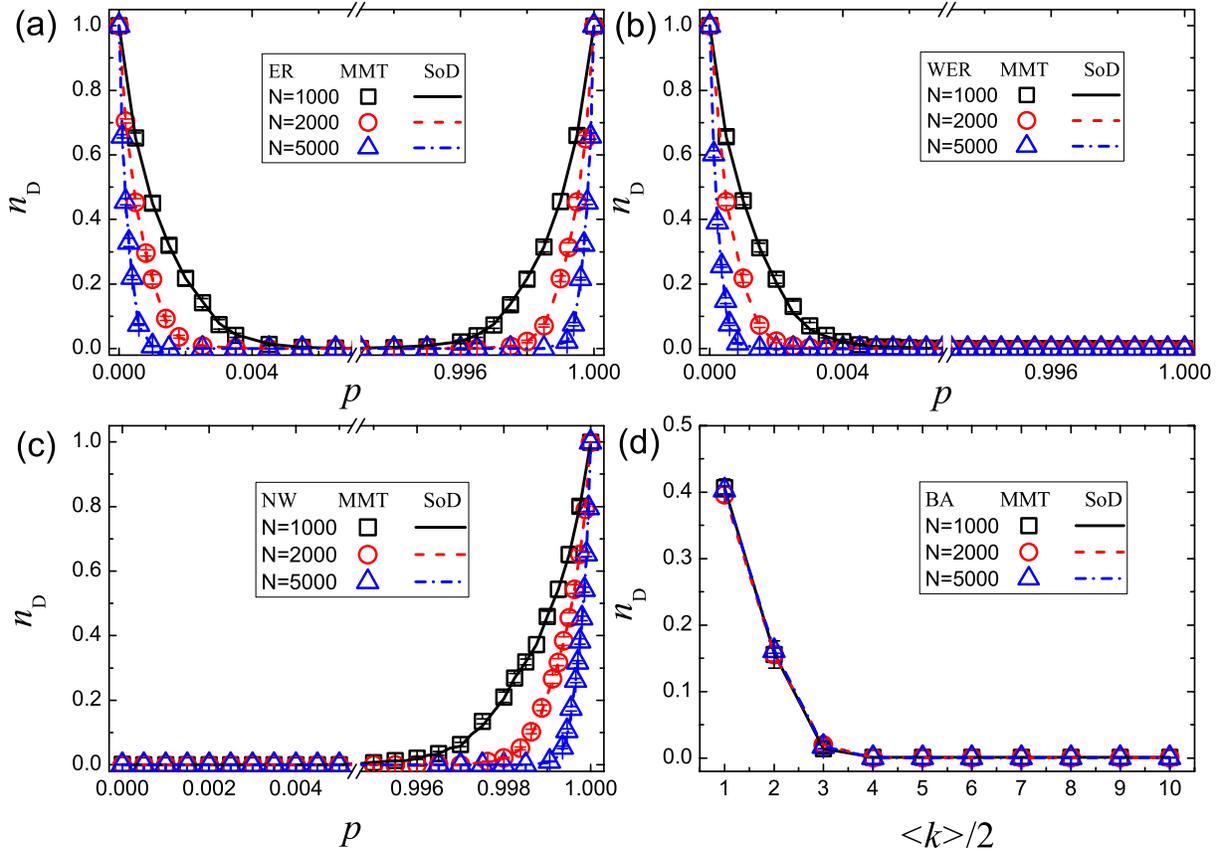,width=\linewidth}
\caption{{\bf Exact controllability of undirected networks}.
Exact controllability measure $n_\text{D}$ as a function of connecting
probability $p$ for (a) unweighted ER random networks and (b) ER random
networks with random weights assigned to links (WER). (c) $n_\text{D}$ versus the probability $p$ of
randomly adding links for NW small-world networks. (d) $n_\text{D}$ versus
half of average degree $\langle k\rangle /2$ for BA scale-free networks. All the networks
are undirected and their coupling matrices are symmetric. The
data points are obtained from the maximum multiplicity theory (MMT) Eq.~(\ref{eq:algebra}) and the
error bars denote the standard deviations, each from 20 independent
realizations. The curves (SoD) are the theoretical predictions of
Eqs.~(\ref{eq:efficient1}) and (\ref{eq:efficient2}) for sparse and dense
networks, respectively. The representative
network sizes used are $N=1000$, 2000 and 5000.}
\label{fig:undirected}
\end{center}
\end{figure}

Figure~\ref{fig:undirected}(a) shows, for
Erd\"{o}s-R\'enyi (ER) random networks~\cite{ER:1960} with identical link
weights, the controllability measure $n_\text{D}$ as a function of the connecting
probability $p$. We see that, regardless of the network size, for small
values of $p$ (e.g., $p<0.006$), $n_\text{D}$ decreases with $p$. However,
for high link density (e.g., $p$'s exceeding 0.994), $n_\text{D}$ begins to
increase toward $(N-1)/N$, which is exact for $p=1$. This counterintuitive
increase in $n_\text{D}$ was not noticed prior to our work, but our exact-controllability
theory reveals that it can be attributed to the impact
of identical link weights. Note that $n_\text{D}$ is symmetric about $p=0.5$,
which can be qualitatively explained by the density function of eigenvalues
of random matrix \cite{Van_Mieghem:book}:
\begin{equation}
\rho(\lambda)\simeq \frac{\sqrt{4Np(1-p)-(\lambda+p)^2}}{2\pi Np(1-p)},
\end{equation}
which is a continuous function. Since the multiplicity measure is discrete,
$\rho(\lambda)$ cannot be used to derive $n_\text{D}$. Nonetheless, the positive
correlation between the peak of $\rho(\lambda)$ and the maximum multiplicity
can be exploited to explain the behavior of $n_\text{D}$. In particular, the maximum
of $\rho(\lambda)$ occurs at $1/[\pi\sqrt{Np(1-p)}]$~\cite{Van_Mieghem:book},
which is symmetric about $p=0.5$, in agreement with the behavior of $n_\text{D}$,
providing an explanation for the symmetry of $n_\text{D}$ about $p=0.5$.
Figure~\ref{fig:undirected}(b) shows $n_\text{D}$ of undirected ER random
networks~\cite{ER:1960} with random link weights. We see
that $n_\text{D}$ is a decreasing function of $p$ and the increasing behavior
in $n_\text{D}$ in the region of high link density disappears.
Figures~\ref{fig:undirected}(c) and (d) show the results
for undirected small-world and scale-free networks in the absence of weights (or
with identical link weights), respectively. We see that the value of $n_\text{D}$ for
Newman-Watts (NW) small-world networks~\cite{NW:1999} is $2/N$ for zero
probability $p$ of randomly adding edges but $n_\text{D}$ quickly reduces to $1/N$
as $p$ increases from zero and remains at the value of $1/N$ until $p$ exceeds
a large value, e.g., 0.994, after which $n_\text{D}$ increases toward $(N-1)/N$. Such
behaviors are analogous to those for the ER random network. The reason that
$n_\text{D}$ is close to the value of $2/N$ for $p=0$ can be attributed to the fact that
the NW small-world networks are constructed from a ring-like network for which
$N_\text{D}=2$ holds. Figure~\ref{fig:undirected}(d) shows, for Barab\'asi-Albert (BA)
scale-free networks~\cite{BA:1999} of different values of the average degree
$\langle k\rangle$, the behavior of $n_\text{D}$. We observe that, for small
values of $\langle k\rangle$, $n_\text{D}$ displays the same decreasing trend as for random networks.

\begin{figure}[H]
\begin{center}
\epsfig{figure=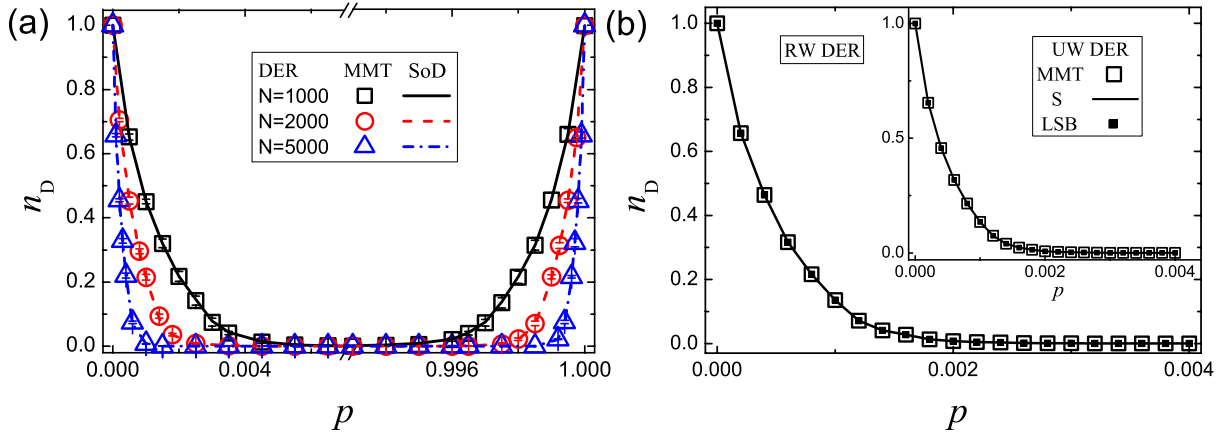,width=\linewidth}
\caption{ {\bf Exact controllability of directed networks.}
(a) Exact controllability measure $n_\text{D}$ as a function of connecting probability $p$
in directed ER random networks with identical weights (DER) for different network
sizes. (b) Exact controllability and structural controllability $n_\text{D}$ versus $p$ in
directed ER random networks with random weights (RW DER) in the absence of bidirectional
links. The inset of (b), Exact controllability and structural controllability
$n_\text{D}$ versus $p$ in unweighted, directed ER random networks (UW DER) in the absence of
bidirectional links. Here, MMT stands for the exact controllability determined by
the maximum geometric multiplicity Eq.~(\ref{eq:geometric}), SoD stands for the
exact controllability of sparse or dense networks from
Eqs.~(\ref{eq:efficient1}) and (\ref{eq:efficient2}), respectively, S denotes the exact
controllability of sparse networks from Eq.~(\ref{eq:efficient1}) and LSB
denotes the structural controllability from the maximum matching algorithm.
In (a), $n_\text{D}^\text{MMT}$ and $n_\text{D}^\text{SoD}$ are in good agreement with each other for different
network sizes. In (b), $n_\text{D}^\text{MMT}$, $n_\text{D}^\text{S}$ and $n_\text{D}^\text{LSB}$ coincide exactly without any difference in the network corresponding to structural matrix. In the inset of (b), the assumption of
structural controllability is weakly violated, ascribed to the unweighted property
of the network. The structural controllability $n_\text{D}^\text{LSB}$ is still
consistent with the exact controllability $n_\text{D}^\text{MMT}$ but with tiny difference. For
example, at $p=8\times 10^{-4}$, $n_\text{D}^\text{MMT}=n_\text{D}^\text{S}=0.21584$, while
$n_\text{D}^\text{LSB}=0.21576$. At $p=10^{-3}$, $n_\text{D}^\text{MMT}=n_\text{D}^\text{S}=0.1352$, whereas
$n_\text{D}^\text{LSB}=0.1350$. For the networks in (a),
the connecting probabilities of two directions between two nodes are both $p$,
but the corresponding random variables are independent of each other. For
the networks in (b), based on the structure of an undirected ER random network, we
randomly assign a direction to each undirected link. All the results are from
20 independent realizations and the error bars denote the standard deviation.
The network size in (b) is 5000.}
\label{fig:directed}
\end{center}
\end{figure}

Figure~\ref{fig:directed} shows the controllability $n_\text{D}$ of directed networks with
identical and random weights, where the latter in the absence of bidirectional
links can be exactly addressed by the structural-controllability framework. In
Fig.~\ref{fig:directed}(a), $n_\text{D}$ of directed ER random networks with identical weights
and possible bidirectional links is somewhat similar to the $n_\text{D}$ of their undirected
counterparts. Thus, high demand of drivers and controllers in dense networks
holds for any network structures with identical link weights. In Fig.~\ref{fig:directed}(b),
a random weight is assigned to each directed link, resulting in a structural
matrix, allowing the comparison between the exact and the structural controllability.
The assumption of structural matrix is ensured by the random weights together with
the directed structure.
We find $n_\text{D}$ resulting from both methods is in complete agreement with each other,
indicating the reduction of the exact-controllability theory to the structural controllability
for structural matrix. We have also tested the difference between the two theories
if the assumption of structural matrix is weakly violated by setting identical
weights, as shown in the inset of Fig.~\ref{fig:directed}(b). We observe consistent
results with tiny difference from each other, suggesting the availability of
structural controllability in directed sparse networks with given link weights and the insensitivity of
controllability to the link weights in sparse networks.

We have also found that the exact controllability of undirected and directed
networks with identical weights can be estimated by the cavity method~\cite{ZOY:2003,ZM:2006}
developed for studying structural controllability, offering a good analytical
prediction and revealing the underlying relationship between the structural
controllability and the exact controllability (Supplementary Fig.~S2 and Note~4).

\begin{table}[P]\small
\caption{{\bf Exact-controllability measures of real unweighted and directed networks}.
For each network, we show its type and name; number of nodes
($N$) and directed links ($L$). $n_\text{D}^\text{MMT}$ is computed from
the maximum geometric multiplicity Eq.~(\ref{eq:geometric}), $n_\text{D}^\text{S}$ is from
Eq.~(\ref{eq:efficient1}), and $n_\text{D}^\text{LSB}$ is the structural-controllability
measure~\cite{LSB:2011}. When random weights are assigned to the originally
unweighted networks, the resulting values of $n_\text{D}^\text{MMT}$ are exactly the
same as those of $n_\text{D}^\text{LSB}$ that can be computed for directed networks with
random link weights. For data sources and references, see Supplementary
Table~S1 and Note~7.}
\begin{center}
\label{tb-2}
\begin{tabular}{ p{3.5cm} p{3.0cm} p{1.5cm} p{1.5cm} p{1.5cm} p{1.5cm} p{1.5cm}}
\hline
\hline
Type  & Name & Nodes & Edges & $n_\text{D}^\text{LSB}$ & $n_\text{D}^\text{MMT}$ & $n_\text{D}^\text{S}$ \\
\hline
Trust  & Prison inmate & 67 & 182 & 0.1343 & 0.1343 & 0.1343 \\

       & WikiVote &	7115 & 103689 & 0.6656 & 0.6656 & 0.6656  \\
 & & & & & & \\		
Food web  & St.Marks & 45 & 224 & 0.3111 & 0.4 & 0.4  \\

       & Seagrass & 49 & 226 & 0.2653 & 0.3265 & 0.3265 \\

       & Grassland & 88 & 137 & 0.5227 & 0.5227 & 0.5227  \\

       & Ythan & 135 & 601 & 0.5111 & 0.5185 & 0.5185  \\

       & Silwood & 154 & 370 & 0.7532 & 0.7662 & 0.7662  \\

       & Little Rock & 183 & 2494 & 0.5410 & 0.7541 & 0.7541  \\
 & & & & & & \\	
Electronic circuits  & s208a & 122 & 189 & 0.2377 & 0.2377 & 0.2377  \\

       & s420a & 252 & 399 & 0.2341 & 0.2341 & 0.2341  \\

       & s838a & 512 & 819 & 0.2324 & 0.2324 & 0.2324   \\
 & & & & & & \\	
Neuronal  & C.elegans  & 297 & 2359 & 0.1650 & 0.1650 & 0.1650  \\
 & & & & & & \\	
Citation  & Small World & 233 & 1988 & 0.6009 & 0.6052 & 0.6052  \\

       & SciMet & 2729 & 10416 &0.4236 & 0.4251 & 0.4251  \\
		
       & Kohonen & 3772 & 12731 & 0.5604 & 0.5620 & 0.5620 \\
 & & & & & & \\		
World Wide Web  & Political blogs & 1224 & 19090 & 0.3562 & 0.3595 & 0.3595  \\
 & & & & & & \\		
Internet  & p2p-1 & 10876 & 39994 & 0.5520 & 0.5531 & 0.5531  \\
	
       & p2p-2 & 8846 & 31839 & 0.5778 & 0.5779 & 0.5779  \\
	
       & p2p-3 & 8717 & 31525 & 0.5774 & 0.5778 & 0.5778  \\
 & & & & & & \\	
Organizational  & Freeman-1 & 34 & 695 & 0.0294 & 0.0294 & 0.0294  \\


       & Consulting & 46 & 879 & 0.0435 & 0.0435 & 0.0435  \\
 & & & & & & \\	
Language  & English Words & 7381 & 46281 & 0.6345 & 0.6345 & 0.6345 \\
	
       & French Words & 8325 & 24295 & 0.6734 & 0.6736 & 0.6736  \\
\hline
\hline
\end{tabular}
\end{center}
\end{table}

Tables~\ref{tb-2} and \ref{tb-3} display, respectively, the exact-controllability measure
of a number of real directed and weighted networks. In both Tables,
$n_\text{D}^\text{MMT}$ and $n_\text{D}^\text{S}$ stand for the result from our exact-controllability
and simplified theory (Eq.~(\ref{eq:efficient1})),
respectively. From Table~\ref{tb-2}, we see that the values of $n_\text{D}^\text{MMT}$ agree
with those of $n_\text{D}^\text{S}$ for all real networks, due to the fact that
these real-world networks are sparse. For Table~\ref{tb-2}, since the networks are
directed, the structural controllability measure, denoted by $n_\text{D}^\text{LSB}$,
can also be calculated. For all cases except the food webs, the values
of $n_\text{D}^\text{MMT}$ and $n_\text{D}^\text{LSB}$ are quite close. In general, we observe the
inequality $n_\text{D}^\text{MMT} \ge n_\text{D}^\text{LSB}$. The consistency between the
results from our exact-controllability theory and from the
structural-controllability theory confirms the similarity between them
for directed and unweighted networks.
Further validation of our theory is obtained by
assigning random weights to all the networks in Table~\ref{tb-2}. In this case,
the coupling matrix of all networks becomes structural matrix and
the new values of $n_\text{D}^\text{MMT}$ are exactly the same as the values of
$n_\text{D}^\text{LSB}$ for the networks associated with structural matrices.
For the originally weighted real-world networks, only the
values of $n_\text{D}^\text{MMT}$ and $n_\text{D}^\text{S}$ are given,
as shown in Table~\ref{tb-3}.
\\

\begin{table}[P]\small
\caption{{\bf Exact-controllability measures of real weighted networks}.
The legends are the same as in Table~\ref{tb-2} except that $n_\text{D}^\text{LSB}$ is
absent. For data sources and references, see Supplementary
Table~S1 and Note~7.}
\begin{center}
\label{tb-3}
\begin{tabular}{ p{4cm} p{4cm} p{1.8cm} p{1.8cm} p{1.8cm} p{1.8cm}  }
\hline
\hline
Type  & Name & Nodes & Edges & $n_\text{D}^\text{MMT}$ & $n_\text{D}^\text{S}$ \\
\hline
Fed Web  & Florida Baydry & 128 & 2137 & 0.25 & 0.25 \\

       & Florida Baywet & 128 & 2106 & 0.2422 & 0.2422 \\

       & Mangrove & 97 & 1492 & 0.2680 & 0.2680 \\
 & & & & & \\

Transportation  & USA top-500 Airport & 500 & 5960 & 0.25 & 0.25 \\
 & & & & & \\
Coauthorships  & Coauthorships & 1461 & 2742 & 0.3436 & 0.3436 \\
 & & & & & \\
Social communication  & Facebook-like & 899 & 142760 & 0.0067 & 0.0011 \\

       & UCIonline & 1899 & 20296 & 0.3239 & 0.3239 \\
 & & & & & \\
Metabolic  & C.elegans & 453 & 2040 & 0.3245 & 0.3245 \\
\hline
\hline
\end{tabular}
\end{center}
\end{table}

\noindent
{\bf \textsf{Discussion}}
\\

We have developed a maximum multiplicity approach to characterize,
exactly, the network controllability in terms of the minimum number of
required controllers and independent driver nodes.
Our approach by transforming the network controllability problem
into an eigenvalue problem, greatly facilitates analysis and offers
a more complete understanding of network controllability in terms of
extensive existing knowledge of network spectral properties. Our theory is
applicable to any network structures, directed or undirected, with or
without link weights and self-loops. Our exact-controllability framework
can reproduce the structural controllability for directed networks associated
with structural matrices. For dense or sparse networks, our theory can
be simplified, resulting in formulas that can be used to calculate
the minimum number of driver nodes in an efficient manner.
This is particularly encouraging, as most real-world complex networks
are sparse. We have studied a large number of real-world networks and
compared the results predicted by the structural-controllability theory
in cases where it is applicable.

Despite the advantage of our theory in studying the controllability
of complex networks, insofar as the exact weights of links are unknown, at the present
the structural-controllability framework is still the best approach to
evaluating controllability of directed networks. This is mainly due to its
error-free characteristic based on the maximum matching algorithm. In contrast, our framework
relies on the eigenvalues and the rank of the network matrix, the evaluation of which inevitably contains
numerical errors, although such error can be controlled by high-accurate
algorithms, such as the SVD algorithm. In addition, the structural-controllability
approach outperforms our method in terms of the computational efficiency
in identifying the minimum set of drivers. For example, in our method, finding
driver nodes based on the elementary transformation requires time on the order of $O(N^2(\log N)^2)$,
whereas the structural-controllability framework requires time on the order of
$O(N^{1/2}L)$ to find unmatched nodes, where $L$ is the total number of links.
Nevertheless, our exact-controllability framework can have broader scope
of application. For example, if the weights of partial links are available,
the framework will offer better measurement of the controllability by setting the
weights of other unavailable links to be random parameters, namely, partial
structural matrix. Our framework
is also valid for undirected networks, where the structural matrix assumption
is slightly violated because of the network symmetry. The framework is as well
applicable to networks containing a number of self-loops with identical or distinct weights,
which will have quite different consequences.
Furthermore, exploring exact controllability is important to achieving
actual control and predicting control energy, especially in man-made networks. Taken together, our
exact-controllability theory as an alternative to the structural-controllability
theory offers deeper understanding of our ability to control complex networked
systems.
\\

\noindent
{\bf \textsf{Method}}
\\

\noindent
{\bf Exact controllability measurement for general networks}
\\

For an arbitrary matrix $A$ in system (\ref{eq:linearSys}),
there exists an $N\times N$ nonsingular matrix $P$ such
that $A=PJP^{-1}$ or
$P^{-1}AP=J$ with $J=\text{diag}(J(\lambda_1),J(\lambda_2),\cdots,J(\lambda_l))$,
where $\lambda_i$ $(i=1,...,l)$ represents the distinct eigenvalues of $A$ and
$J(\lambda_i)$ is the Jordan block matrix of $A$ associated with $\lambda_i$
\cite{HJ:book}. For every distinct eigenvalue $\lambda_i$ of matrix $A$, there are
two concepts of multiplicity~\cite{HJ:book}.
(i) Algebraic multiplicity $\delta(\lambda_i)$: multiplicity of
eigenvalue $\lambda_i$, which satisfies $\sum_{i=1}^{l}\delta(\lambda_i)=N$, where
$\delta(\lambda_i)$ is also the eigenvalue degeneracy;
(ii) Geometric multiplicity $\mu(\lambda_i)$: the dimension of the eigenspace
of $A$ associated with $\lambda_i$, i.e.,
$\mu(\lambda_i)=\dim V_{\lambda_i}=N-\text{rank}(\lambda_i I_N-A)$.
Note that geometric multiplicity is determined by the eigenvectors associated
with the eigenvalues, which means that both multiplicities depend on the eigenvalues
of the matrix $A$. In general, we have $\mu(\lambda_i)\le \delta(\lambda_i)$. The
two types of multiplicities will be
used to derive our theory of exact controllability. According to the
definition of $\mu(\lambda_i)$, we can find that $\mu(\lambda_i)$ is equal to the
number of basic Jordan blocks in $J(\lambda_i)$~\cite{HJ:book}, that is,
$J(\lambda_i)=\text{diag}(j_1,j_2,\cdots,j_{\mu(\lambda_i)})$, where
$j_s$ $(s=1,2,\cdots,\mu(\lambda_i))$ is the basic Jordan-block matrix having the
form exemplified in Eq.~(\ref{eq:jordan}) below. The key to our theory lies in bridging
the two multiplicities and the controllability via the PBH rank condition~\cite{PBH}.

Applying the nonsingular transformation $\mathbf{y}=P^{-1}\mathbf{x}$ and $Q=P^{-1}B$,
we can rewrite system (\ref{eq:linearSys}) in Jordan form as
\begin{equation} \label{eq:nonsinglar}
\dot{\mathbf{y}}=J\mathbf{y}+Q\mathbf{u}.
\end{equation}
It can be verified that systems (\ref{eq:linearSys}) and (\ref{eq:nonsinglar}) have
the same controllability in the sense that
$\text{rank}(\lambda I_N-A,B)=\text{rank}(\lambda I_N-J,Q)$,
$\forall \lambda\in \sigma(A)$ with $\text{rank}(B)=\text{rank}(Q)$. We call $Q$ the
transformed matrix, which can be used to calculate the control energy
(Supplementary Note~5).

Any basic Jordan block matrix $j(\lambda)$ associated with eigenvalue $\lambda$ has the form:
\begin{equation} \label{eq:jordan}
j(\lambda)= \begin{pmatrix}
\lambda&1&0&\cdots&0&0\\
0&\lambda&1&\cdots&0&0\\
0&0&\lambda&\cdots&0&0\\
\vdots&\vdots&\vdots& &\vdots&\vdots\\
0&0&0&\cdots&\lambda&1\\
0&0&0&\cdots&0&\lambda\\
\end{pmatrix}.
\end{equation}
For each $j(\lambda)$ of order $v$, all elements in the first column of matrix
$\lambda I_v-j(\lambda)$ are zero and other $v-1$ columns are independent of each
other. We thus have $\text{rank}(\lambda I_v-j(\lambda))=v-1$. Since the number
of basic Jordan blocks associated with eigenvalue $\lambda_i$ is equal to its
geometric multiplicity $\mu(\lambda_i)$, the number of zero columns in
$\lambda_i I_N-j(\lambda)$ is $\mu(\lambda_i)$ and other
$N-\mu(\lambda_i)$ columns are independent of each other, yielding
$\text{rank}(\lambda_i I_N-J)=N-\mu(\lambda_i)$.

The PBH rank condition~\cite{PBH} stipulates that system (\ref{eq:linearSys}) is
controllable if and only if for each $\lambda_i$ of matrix $A$, the following holds:
\begin{equation} \label{eq:rank_ori}
N=\text{rank}(\lambda_i I_N-A,B)=\text{rank}(\lambda_i I_N-J,Q).
\end{equation}
Due to the rank inequality~\cite{Strang:book}, Eq.~(\ref{eq:rank_ori}) can be rewritten
as $N=\text{rank}(\lambda_i I_N-J,Q) \le \text{rank}(\lambda_i I_N-J)+\text{rank}(Q)$,
so that $\text{rank}(Q) \geq N-\text{rank}(\lambda_i I_N -J)$. Recalling that
$\text{rank}(Q)=\text{rank}(B)$, we have
\begin{equation} \label{eq:rankB}
\text{rank}(B) \geq N - \text{rank}(\lambda_i I_N - J).
\end{equation}
Equation~(\ref{eq:rankB}) will be satisfied if $\text{rank}(B)$ is larger than or
equal to the maximum value of $N-\text{rank}(\lambda_i I_N - J)$ for all eigenvalues
$\lambda_i$. It can be shown that the equality in Eq.~(\ref{eq:rankB}) can be satisfied
by specific construction of $Q$, indicating the necessity of establishing $Q$ via
nonsingular transformation associated with Jordan blocks (Supplementary Note~2).
In this regard, the minimum value of $\text{rank}(B)$ is the maximum value of
$N-\text{rank}(\lambda_i I_N - J)$. That is, the minimum number $N_\text{D}$ of independent
driver nodes, $\min \{\text{rank}(B)\}$, is $\max_i\{ N-\text{rank}(\lambda_i I_N-J)\}$.
Since this is equivalent to $\max_i\{N-\text{rank}(\lambda_i I_N-A)\}$, according to
the definition of geometric multiplicity, we have Eq.~(\ref{eq:geometric}): $N_\text{D}=\max_i\{\mu(\lambda_i)\}$.
Although the maximum geometric multiplicity as the lower bound of the number of controllers
has been implied in linear control theory~\cite{AM:book,Sontag:book} and some literature~\cite{LFsys,PN:2012,NP:2012},
we prove that the maximum multiplicity is both the necessary and sufficient conditions to ensure full control
and offer a rigorous scheme to identify the minimum set of drivers, leading to the development
of an exact and efficient controllability framework.
If $A$ is diagonalizable, e.g., for undirected networks, then we have
$\mu(\lambda_i)=\delta(\lambda_i)$ for each eigenvalue and the number of independent
drivers or controllers under this condition is determined by Eq.~(\ref{eq:algebra}):
$N_\text{D}=\max_i\{\delta(\lambda_i)\}$.

Our theory thus offers an exact characterization of controllability of complex networks
with any structures and weights in the presence or absence of self-loops.
It should be emphasized that $N_\text{D}=\max_i{\{\delta(\lambda_i)\}}$
is valid for undirected network or diagonalizable matrix $A$. In the latter case, we have
$J=\Lambda=\text{diag}(\lambda_1,\lambda_2,...,\lambda_N)$ ($\lambda_i$ for different
$i$ can be of the same value). In addition, we can show that the observability of
complex networks can be formulated in a similar way (Supplementary Note~6).
\\

\noindent
{\bf Exact controllability for dense and sparse networks}
\\

For a large sparse network in the absence
of self-loops, the expectation of eigenvalues of the coupling matrix $A$ is
$E(\lambda) = \frac{1}{N}\sum_{i=1}^N \lambda_i = \frac{1}{N}\sum_{i=1}^N a_{ii}=0$, so
the maximum geometric multiplicity occurs at eigenvalue $\lambda=0$ with high
probability~\cite{Papoulis:book}. Moreover, it can be proven that the geometric
multiplicity associated with the zero eigenvalue is equal to the rank deficiency
\cite{Strang:book}: $\mu(0) = N- \text{rank}(A)$. The minimum number of controllers and
independent drivers as determined by the geometric multiplicity associated with the zero
eigenvalue can then be estimated by Eq.~(\ref{eq:efficient1}):
$N_\text{D} = \max\{1,N-\text{rank}(A)\}$.

For a densely connected network with unit link weights, the zeros in $A$ scales with $N$ in the limit of large
$N$. In the absence of loops, the probability is high to find two groups of nodes that
are interconnected and have mutual neighbors. In general, in this case the coupling
matrix consists of many rows of the form
\begin{equation} \label{eq:-1matrix}
\left[
\begin{array}{cccccc}
0&1&a_{13}&a_{14}&\cdots&a_{1N}\\
1&0&a_{13}&a_{14}&\cdots&a_{1N}\\
\cdots&\cdots&\cdots&\cdots&\cdots&\cdots
\end{array}
\right].
\end{equation}
The corresponding rows in $\text{det}(\lambda I_N-A)$ of the characteristic polynomial are
\begin{equation}
\begin{vmatrix}
\lambda&-1&-a_{13}&-a_{14}&\cdots&-a_{1N}\\
-1&\lambda&-a_{13}&-a_{14}&\cdots&-a_{1N}\\
\cdots&\cdots&\cdots&\cdots&\cdots&\cdots
\end{vmatrix}.
\end{equation}
We see that if $\lambda =-1$, the two rows are identical. We then have~\cite{Van_Mieghem:book},
$\text{det}(-I_N-A)=\text{det}(I_N+A)=0$. As a result, $-1$ is the eigenvalue of matrix
(\ref{eq:-1matrix}). Since, in a dense network with unit weights, the likelihood to observe two rows with
the form in matrix (\ref{eq:-1matrix}) is high, $\lambda =-1$ becomes the eigenvalue
corresponding to maximum multiplicity,  enabling an efficient formulation of exact
controllability for dense networks. In particular, the geometric multiplicity is defined
as $N-\text{rank}(\lambda_i I_N - A)$. Substituting $\lambda$ by $-1$ yields
$N_\text{D} = \max\{1,N-\text{rank}(I_N+A)\}$. For identical link weights with arbitrary value $w$,
the eigenvalue becomes $-w$ rather than $-1$. Substituting $\lambda$ by $-w$ yields
Eq.~(\ref{eq:efficient2}): $N_\text{D} = \max\{1,N-\text{rank}(wI_N+A)\}$.
Equations~(\ref{eq:efficient1}) and (\ref{eq:efficient2}) are the exact-controllability
theory for any sparse networks and dense networks with identical link weights, respectively.

It is noteworthy that, although we assume the absence of self-loops so as to
derive the simplified formulas (\ref{eq:efficient1}) and (\ref{eq:efficient2}), they are
still valid in the presence of a small fraction of self-loops, insofar as the self-loops
do not violate the dominations of 0 and $-w$ in the eigenvalue spectrum of matrix $A$ for
sparse and dense networks, respectively. In fact, the combination of the number and the
weight distribution of self-loops plays quite intricate roles in the controllability,
which will be discussed thoroughly elsewhere.
\\

\noindent
{\bf Computation of matrix rank and eigenvalues}
\\

We use the SVD method~\cite{SVD} to compute the eigenvalues of matrix $A$ of any complex networks
required by the general maximum multiplicity theory for quantifying the exact controllability
of complex networks. To count the number of eigenvalues with the same value, that is, the
algebraic multiplicity, we set a small threshold $10^{-8}$. If the absolute difference between
two eigenvalues is less than the threshold, they are regarded as identical. We have checked
that the maximum algebraic multiplicity is very robust to the setting of the threshold.
This is due to the fact that this multiplicity usually arises at integer
eigenvalues that are independent of the threshold.

The SVD method can also be used to calculate the accurate rank of matrix associated with an
arbitrary network. For sparse networks and dense networks with identical weights, the exact
controllability is only determined by the rank of matrix. For sparse, undirected networks with
identical or random weights, we have
checked that the LU decomposition method~\cite{LU} can yield reliable results of the matrix rank, in good
agreement with those from the SVD method. However, for other types of networks, we have to
rely on the SVD method. For both LU and SVD methods,
the tolerance level is set to be $10^{-8}$. If a value in the diagonal is less than the
threshold, it will be treated as zero to determine the matrix rank. We have examined
that the matrix rank is insensitive to the threshold insofar as it is sufficiently small.
All the exact controllability measures of real networks
presented in Table II and III are obtained via the SVD method to ensure accuracy.
The structural controllability of networks is computed via the maximum matching algorithm~\cite{LSB:2011}.

For the elementary column transformation for identifying driver nodes, we use
the Gaussian elimination method with computation complexity
$O(N^2(\log N)^2)$~\cite{GaussEle} to yield the column
canonical form that reveals the linear dependence among rows.

\section*{Acknowledgements}
We thank Mr. X. Yan, Prof. Y. Fan, Prof. S. Havlin, Dr. B. Ao
and Prof. L. Huang for valuable comments and discussions.
Z.Y. was supported by STEF under Grant No. JA12210.
W.-X.W. was supported by NSFC under Grant No. 11105011, CNNSF 
under Grant No.61074116 and
the Fundamental Research Funds for the Central Universities.
Y.-C.L. was supported by AFOSR under Grant No.
FA9550-10-1-0083, and by NSF under Grants No. CDI-1026710 and No. BECS-1023101.

\section*{Author contributions}
W.-X.W., Z.Y., Z.D., and Y.-C.L. designed research; Z.Y. and C.Z. performed research;
Z.Y. and W.-X.W. contributed analytic tools; W.-X.W., C.Z, Z.D., and Y.-C.L. analyzed
data; and W.-X.W. and Y.-C.L. wrote the paper.

\section*{Additional information}
The authors declare no competing financial interests.

\end{document}